\documentclass[aps,prl,twocolumn,showpacs,superscriptaddress,groupedaddress]{revtex4}  
\usepackage{graphicx}  
\usepackage{dcolumn}   
\usepackage{bm}        
\usepackage{amssymb}   
\usepackage{color}
\usepackage{amsmath,amssymb}

\def \beq {\begin{equation}}
\def \edq {\end{equation}}
\def \bes {\begin{subequations}}
\def \eds {\end{subequations}}
\def \beqn {\begin{equation*}}
\def \edqn {\end{equation*}}

\def \dag {\dagger}
\def \up {\uparrow}
\def \down {\downarrow}

\def \veps {\varepsilon}
\def \tveps {\tilde{\varepsilon}}
\def \tGamma {\tilde{\Gamma}}
\def \sm {\sigma}

\def \calh {{\cal{H}}}
\def \calt {{\cal{T}}}

\begin{document}
\title{A thermally driven out-of-equilibrium two-impurity Kondo system}
\author{Miguel A. Sierra}
\affiliation{Institut de F\'{\i}sica Interdisciplin\`aria i de Sistemes Complexos
IFISC (CSIC-UIB), E-07122 Palma de Mallorca, Spain}
\author{Rosa L\'{o}pez}
\affiliation{Institut de F\'{\i}sica Interdisciplin\`aria i de Sistemes Complexos
IFISC (CSIC-UIB), E-07122 Palma de Mallorca, Spain}
\author{Jong Soo Lim}
\affiliation{School of Physics, Korea Institute for Advanced Study, Seoul 130-722, Korea}

\begin{abstract}
The archetypal two-impurity Kondo problem in a serially-coupled double quantum dot is investigated in the presence of a thermal bias $\theta$.
The slave-boson formulation is employed to obtain the nonlinear thermal and thermoelectrical responses. 
When the Kondo correlations prevail over the antiferromagnetic coupling $J$ between dot spins we demonstrate that the setup shows negative differential thermal conductance regions behaving as a thermal diode. Besides, 
we report a sign reversal of the thermoelectric current $I(\theta)$ controlled by $t/\Gamma$ ($t$ and $\Gamma$ denote the interdot tunnel and reservoir-dot tunnel couplings, respectively) and $\theta$. All these features are attributed to the fact that 
at large $\theta$, both $Q(\theta)$ (heat current) and $I(\theta)$  are suppressed regardless the value of $t/\Gamma$ because the double dot decouples at high thermal biases. Eventually, and for a finite $J$,  we investigate how the Kondo-to-antiferromagnetic crossover is altered by $\theta$. 
\end{abstract}
\maketitle
\emph{Introduction---}The richness of the single impurity Anderson model has been nicely exhibited in singly-occupied quantum dots (QDs) attached to electronic reservoirs, in which the Kondo effect is its most prominent feature \cite{Ng88,Gla88,Kaw91}. 
Such artificial systems allow us an unprecedented benefit to investigate out-of-equilibrium many-body effects with the additional intriguing possibility of tuning the parameters controlling its physics \cite{Dgo98,Cro98,Sch98,Sas00}. 
The Kondo effect is built from the antiferromagnetic (AF) correlations between the delocalized electrons at the reservoirs and the localized spin in the QD \cite{hew93}. 
It is reflected in the differential conductance as a peak
of height $2e^2/h $ \cite{Wie00} and width $\approx k_BT_K$ at zero applied voltage called the zero bias anomaly (ZBA) ($k_B=1$ is the Boltzmann constant and $T_K$ corresponds to the Kondo temperature). 
Importantly, the recent experimental boost on quantum transport through correlated quantum systems has opened a new arena:  the out-of-equilibrium Kondo physics in more intricate nanostructures such as artificial coupled Kondo systems \cite{Jeo01,Cra04,Wah07,Nee11,Chor12,Lar13}. The paradigmatic two-impurity Kondo system (2IKS) \cite{And64,Jay81,Jon87,Jon88,Jon89a,Jon89b,Jon91} possesses a quantum phase transition (QPT) in which the critical point is a two-channel Kondo fixed point \cite{Aff92,Jun95,Zar06,Fre11} recently observed in QDs \cite{Pot06,Kel15}. 
Such QPT is, indeed, a consequence of the competition between the Kondo effect and the spin-to-spin interaction, the latter attributed to the Ruderman-Kittel-Kasuya-Yosida interaction \cite{Pas05,Min10} (RKKY) which couples two localized spins through its interaction with the delocalized electrons.  
Therefore, for an antiferromagnetic exchange coupling $J>0$ the spins exist in a singlet state \cite{Sak90,Geo99,Bus00,Aon01}. 
In a Kondo dominant regime $T_K\gg J$, each localized spin exhibits a Kondo singularity, whereas for $J\gg T_K$ the spins are locked into a singlet state with suppressed Kondo correlations. 
In the context of  artificial Kondo molecules, the antiferromagnetic interaction is generated via superexchange mechanism with $J\approx 4t^2/U$ with $t$ the interdot tunneling coupling and $U$ the interdot Coulomb repulsion strength. Now, 
the connection between both reservoirs transforms the QPT into a crossover  \cite{Geo99,Bus00,Aon01,Kon07,Pas05b,Vav05,Bin02,Ham13}. 
Different theoretical techniques, like Numerical Renormalization Group \cite{Bul08,Rok09,Sak90,Zit07} and others \cite{Aon01,Bus00,Kar06}  have been applied to a deeper understanding  of the 2IKS, 
most of them are applicable at equilibrium and only a few focused on the nonequilibrium configuration \cite{Ram00,Ros02,Bin02,Pas05,Kon07,Vav05,Ham13}.  Experimentally, the behavior of the 2IKS has been probed in tunnel-coupled QDs \cite{Jeo01,Cra04,Wah07,Nee11,Chor12,Lar13}.  

\begin{figure}[t!]
\centering
\includegraphics[width=0.45\textwidth]{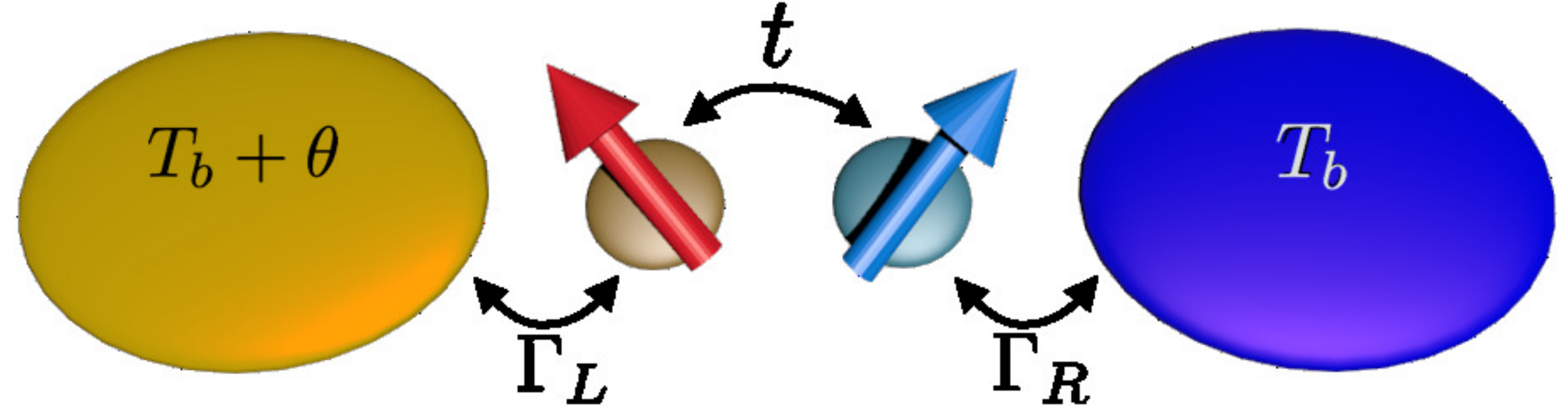}
\caption{Schematic of a tunnel-coupled double quantum dot system. Each reservoir is coupled to its respective dot with the tunnel hybridization $\Gamma_{L/R}$ and the dots are coupled with a tunnel amplitude $t$. The left reservoir is heated meanwhile the right reservoir remains at the background temperature $T_b$ giving rise to a temperature gradient $\theta$ in the system.}
\label{fig:0}
\end{figure}
However, out-of-equilibrium physics in the 2IKS has been analyzed only with electric fields. A
much less investigated situation is when a thermal gradient  $\theta=T_L-T_R$ ($T_{L/R}$ is the temperature of the left/right reservoir) is applied.  Strong thermoelectrical response has been demonstrated to occur in confined nanostructures such as QDs \cite{mol105,dzu101,sfs105,xch168,pro165,hth123}. These systems allow to partially convert electricity into heat and viceversa with large Seebeck coefficients \cite{handbook}. In particular, the case of a thermally biased QD in the Kondo regime has been investigated recently, mostly in the linear regime \cite{zzh086}. In Ref.~\cite{Mig17} it was shown that Kondo correlations are destroyed with $\theta$, but more slowly than with the applied voltage $V_{\rm sd}$. 
Besides,  it was found that the thermoelectric current $I(\theta)$ exhibits a nontrivial zero in the Kondo regime at a critical thermal bias  $\theta_c\approx T_K$.  Consequently, the sign of $I(\theta)$ can be controlled externally by varying $\theta$. 
A sign reversal of $I(\theta)$ is a very inusual phenomenon that has been recently reported to happen in single quantum dots in the Coulomb blockade regime \cite{Mig14} and in tubular nanowires in the presence of a magnetic field as its sign control parameter \cite{david}.

In this work, we focus on the nonlinear response of a thermally-biased  2IKS (see Fig.~\ref{fig:0}). 
Remarkably, with $T_K\gg J$, we find that a thermal bias perturbs our serial DQD differently depending on the ratio between the dot-lead ($\Gamma$) and interdot tunneling ($t$) couplings. We report situations where the DQD setup operates as a thermal diode \cite{Tdiode} exhibiting a region of negative differential thermal conductance for sufficiently large temperature biases. Thermal diodes are of great importance for their potential applications as coolers \cite{refrigeration}, energy harvesters \cite{reviewthermal} or  thermal memory storage \cite{Wang08}, and in the emergent field of coherent caloritronics  \cite{coherentcal}.

 In contrast when $J > T_K$, the dependence of $T_K$ for each dot on $\theta$ has a strong impact on the Kondo-to-antiferromagnetic crossover. 

\emph{Model Hamiltonian and theoretical approach---}
Our system illustrated in Fig.~\ref{fig:0} consists of a tunnel-coupled double quantum dot where each dot is connected to a electronic reservoir acting as an artificial Kondo molecule. 
The Hamiltonian describing this system reads
\beq
\label{eq:MHam}
\calh = \calh_C + \calh_D + \calh_T\,.
\edq 
$\calh_C = \sum_{\alpha,k,\sm} \veps_{\alpha k} c^{\dag}_{\alpha k\sm} c_{\alpha k\sm}$
describes the fermionic reservoirs.
$c^{\dag}_{\alpha k\sm}$ ($c_{\alpha k\sm}$) creates (annihilates) an electron of energy $\veps_{\alpha k}$ with wavenumber $k$ 
and spin $\sm=\{\up,\down\}$ in the reservoir $\alpha=\{L,R\}$.
The dot Hamiltonian is expressed in the form 
$\calh_D = \sum_{\alpha,\sm} \veps_{\alpha} d^{\dag}_{\alpha\sm} d_{\alpha\sm} + U n_{\alpha\up} n_{\alpha\down}+ \sum_{\sm} t d_{L\sm}^{\dag}  d_{R\sm}+h.c.$
$d^{\dag}_{\alpha\sm} (d_{\alpha\sm})$ creates (annihilates) an electron of energy $\veps_{\alpha}$ with spin $\sm$ in the dot $\alpha$ 
and $n_{\alpha\sigma}=d_{\alpha\sigma}^\dagger d_{\alpha\sigma}$ denotes the dot occupation operator. Electrons hop between dots with a tunnel amplitude $t$. Two electrons located in the same dot feel the on-site Coulomb interaction $U$.
$\calh_T = \sum_{\alpha k\sm} V_{\alpha} c^{\dag}_{\alpha k\sm} d_{\alpha\sm} + h.c$ connects
each dot to its respective reservoir with tunneling amplitude $V_{\alpha}$.

To observe thermal and thermoelectric effects, we raise the temperature of the left reservoir by an amount of $\theta>0$ with respect to the background temperature such that
$T_L = T_b + \theta$ and $T_R = T_b$. 
With this configuration, our analysis proceeds in two directions: (i) the Kondo dominant case with  a negligible $J\approx 0$ and (ii)  the crossover from Kondo-to-AF phases for $J \ne 0$. 

We consider the infinite-$U$ limit, thereby each dot is singly occupied for $\varepsilon_\alpha <0$. Then, at very low temperatures, we can adopt a  Fermi liquid description \cite{Col87}. In such case
the Hamiltonian is reformulated employing the slave-boson representation. The dot operators become $d_{\alpha \sigma}^{\dag}=f_{\alpha \sigma}^{\dag}b_{\alpha}$.
The bosonic operator $b_{\alpha}$ annihilates  an  empty  state, while the auxiliary fermion  operator $f_{\alpha \sigma}^{\dag}$ creates the singly occupied state with spin $\sigma$ at the  dot $\alpha$. 
The single occupancy constraint $ b_{\alpha}^{\dag}b_{\alpha} + \sum_\sigma f_{\alpha \sigma}^{\dag}f_{\alpha \sigma}=1$ is enforced by introducing Lagrange multipliers $\Lambda_{\alpha}$. 
We apply a $1/N$ expansion, with $N$ as the level degeneration of the dot angular momentum, here $N=2$, and keep the leading order. Then, the bosonic operators are replaced by their mean-field  (MF) values $\langle b_{\alpha}\rangle \to\sqrt{N}\tilde{b}_{\alpha}$
neglecting the fluctuations around its expectation value. 
The MF approach has been successfully applied to a number of Kondo-related problems \cite{Col87,Jon89b,Geo99,Aon01} such as double-dot systems driven out of equilibrium \cite{Ram00,Ros02,Bin02,Pas05}. We obtain a set of MF equations, i.e., the boson equation of motion  and the Lagrange condition. By solving these equations, we determine the Kondo impurity parameters ($\Lambda_\alpha$, $\tilde{b}_{\alpha}$) that renormalize the system parameters as: $\tveps_{\alpha}= \veps_{\alpha}+\Lambda_\alpha$ and $\tGamma_{\alpha} = \Gamma_{\alpha}|\tilde{b}_{\alpha}|^2$ with $\Gamma_{\alpha} = \pi\rho_{\alpha} |V_{\alpha}|^2$ and
$\rho_{\alpha}$ as the density of states (DOS) of the reservoir $\alpha$ considered constant in a bandwidth $D$.  Besides, $\tGamma_{L/R}$, and $\tveps_{L/R}$ represent the effective Kondo temperatures $T_{KL/R}$ and  the Kondo resonance positions within this approach. 
The details of the calculation of $\tveps_{\alpha}$ and $\tGamma_{\alpha}$ can be found in the supplementary material \cite{SM}.

The electric and heat current flowing out of the left reservoir ($I=I_L$ and $Q=Q_L$) are obtained from the formula $\mathcal{H}$ giving the result
\begin{eqnarray}
I &=& \frac{e}{h}\int d\omega~\left[f_L(\omega) - f_R(\omega)\right] \calt(\omega) \,,\\
Q &=& \frac{1}{h}\int d\omega~\left[f_L(\omega) - f_R(\omega)\right](\omega-\mu_\alpha) \calt(\omega) \,,
\end{eqnarray} 
where $e>0$. The Fermi-Dirac distribution of the reservoir $\alpha$ with the chemical potential $\mu_{\alpha}$
is denoted as $f_\alpha(\omega)=1/\left\{1+\exp\left[(\omega-\mu_{\alpha})/k_BT_{\alpha}\right]\right\}$. The analytical expressions for the transmission $\calt(\omega)$ and derivations for charge and heat flows are given in the Supplementary information \cite{SM}. We emphasize that such currents have a nontrivial dependence on $\theta$ through the MF parameters.

\emph{Results (I): Thermally driven Kondo regime---}

\begin{figure}
\centering
\includegraphics[width=0.45\textwidth]{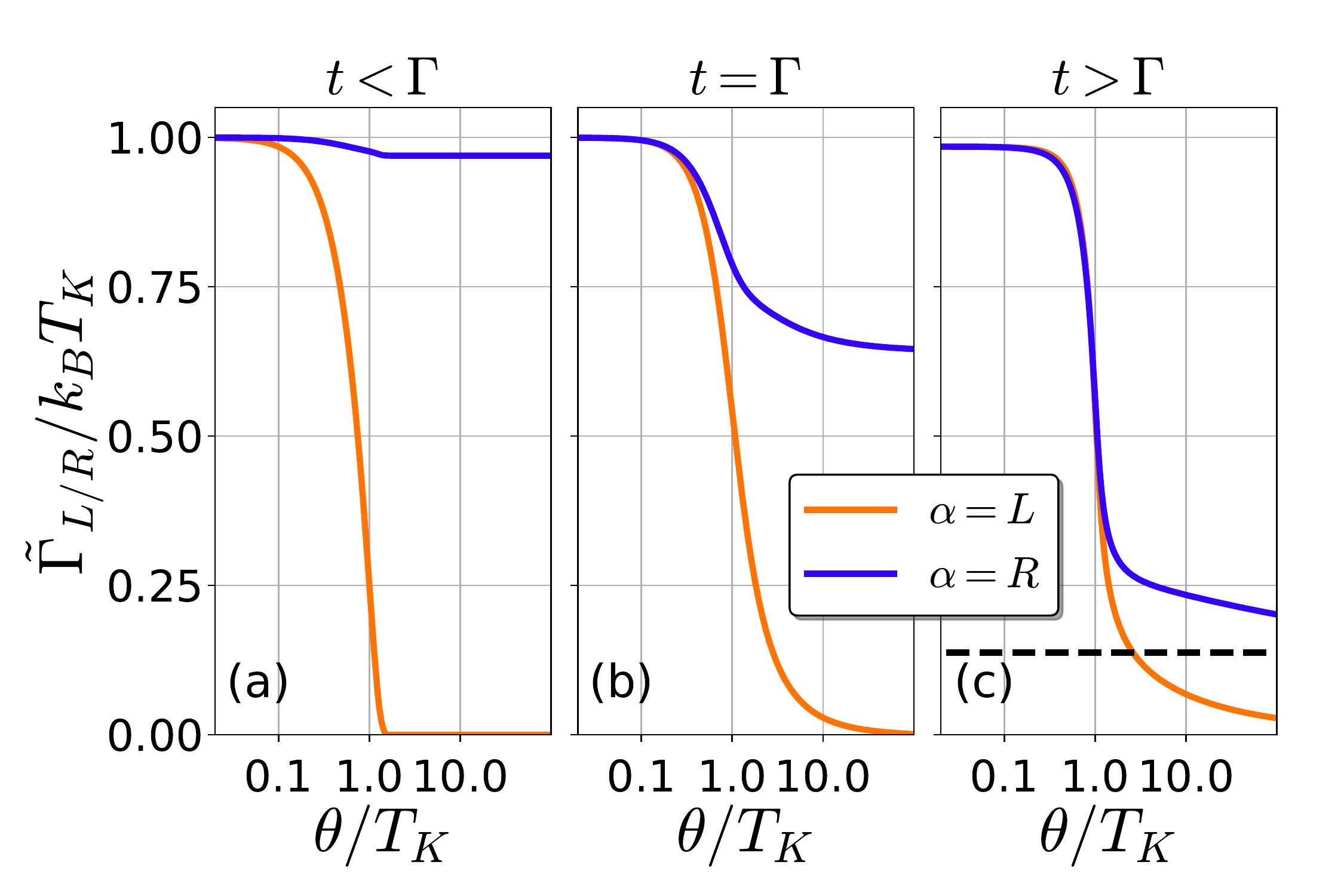}
\caption{$\tGamma_{L/R}$  as a function of the thermal bias $\theta$. 
The left, middle, and right columns correspond to the weak ($t=0.25\Gamma$), intermediate ($t=\Gamma$), and  the strong coupling regimes ($t=2.5\Gamma$), respectively. 
The parameters are normalized by the common Kondo temperature $T_K$ [for the definition of $T_K$, see the main text]. 
Parameters: $k_B=1$, $\Gamma_L=\Gamma_R=\Gamma$, $\veps_L=\veps_R=-3.5\Gamma$, $T_b=10^{-5}\Gamma$, $D=100\Gamma$, $\mu_{L,R}=\varepsilon_F=0$.}
\label{fig:1}
\end{figure}
At this stage we focus on a negligible $J\approx 0$ value.  Hereafter, we consider a symmetric double-dot system: $\veps_0\equiv\varepsilon_{L}=\varepsilon_{R}=-3.5\Gamma$ and $\Gamma_L=\Gamma_R=\Gamma$.
We set $D=100\Gamma$ with a background temperature  $T_b=10^{-5}\Gamma$, thereby $T_b\ll T_{K0}$ where $T_{K0}=D\exp\left[\pi\veps_0/\Gamma\right]$ denotes the Kondo temperature for a single dot \cite{hew93}, where $T_{K0}=0.0016\Gamma$ for the chosen bare system parameters. For a serial DQD, the characteristic Kondo temperature is normalized by the tunneling amplitude accordingly to $T_K = T_{K0}\exp\left[(t/\Gamma)\tan^{-1}(t/\Gamma)\right]/\sqrt{1+(t/\Gamma)^2}$ \cite{Aon01}. Firstly, we explore in 
Fig.~\ref{fig:1} the behavior of the MF parameters $\tGamma_{\alpha}$ and $\tveps_{\alpha}$ when a thermal gradient $\theta$ is applied. We distinguish three different scenarios, namely (i) the weakly coupling regime $t/\Gamma<1$, (ii) the intermediate $t/\Gamma\sim 1$, and (iii) the strong coupling regime $t/\Gamma>1$ regime. Then, when $t/\Gamma<1$ [see Fig.~\ref{fig:1}(a)]
we observe that  $\tGamma_{L}(\theta)$ and $\tGamma_{R}(\theta)$ behave very differently with $\theta$. Here,   
$\tGamma_{L}$  (hot reservoir) decreases rapidly as $\theta$ increases, whereas  $\tGamma_{R}$ remains almost unaffected by $\theta$. Each dot develops an independent Kondo resonance with its own reservoir that yields a much lower Kondo temperature for the hot reservoir that decreases its value as long as $\theta$ augments. 
However, when $t/\Gamma \approx 1$, both $\tGamma_{L}$ and $\tGamma_{R}$ are affected as shown in Fig.~\ref{fig:1}(b) due to the moderate coupling between the Kondo resonances. Eventually for $t/\Gamma > 1$  the DQD behaves effectively as a Kondo coherent molecule exhibiting bonding and antibonding Kondo states. 
Here, $\tGamma_{L}$ and $\tGamma_{R}$ are similar when $\theta$ is tuned. 
However, at large $\theta$, Kondo correlations disappear in the hotter side. 
In such case, the right dot decouples completely from the left dot and $\tGamma_{R}$ approaches $T_{K0}$ as indicated by the dashed line in Fig.~\ref{fig:1}(c). 
Consequently, for very large $\theta$ values the transport through the double dot is blocked even though Kondo correlations in the right 
(cold) dot take place.  

\begin{figure}
\centering
\includegraphics[width=0.51\textwidth]{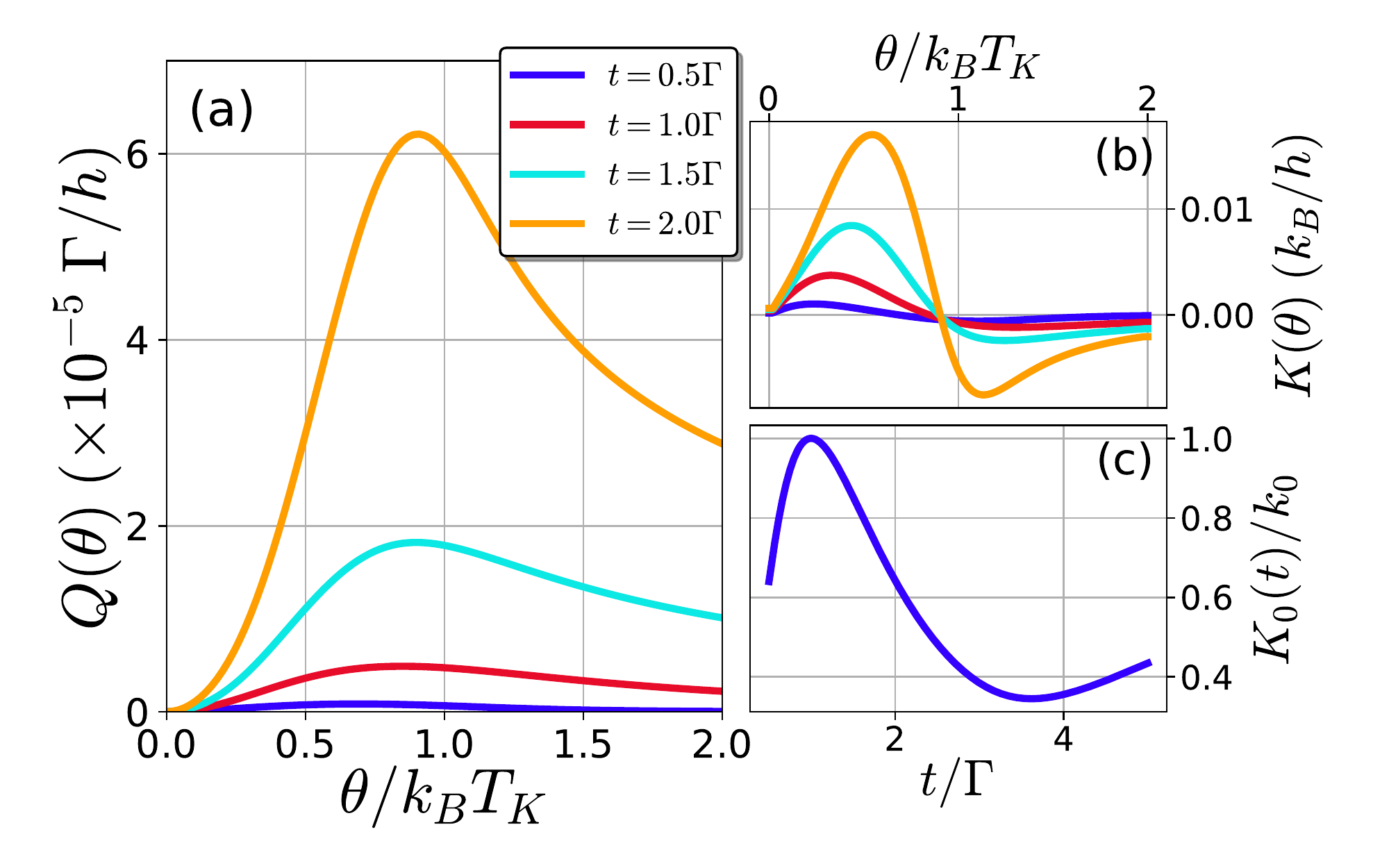}
\caption{(a) Heat current $Q(\theta)$ for different values of the tunnel coupling $t$. (b) Differential thermal conductance $K(\theta) = dQ/d\theta$ for different values of the tunnel coupling $t$. (c) Linear thermal conductance $K_0(t)$ normalized with the thermal conductance quantum $k_0$ as a function of the tunnel amplitude $t$. This result is equivalent to the transmission function at the Fermi level $K_0(t) = k_0\mathcal{T}(\omega=0,t)$. Same parameters as in Fig.~\ref{fig:1}.} 
\label{fig:2}
\end{figure}

\begin{figure}[t]
\centering
\includegraphics[width=0.45\textwidth]{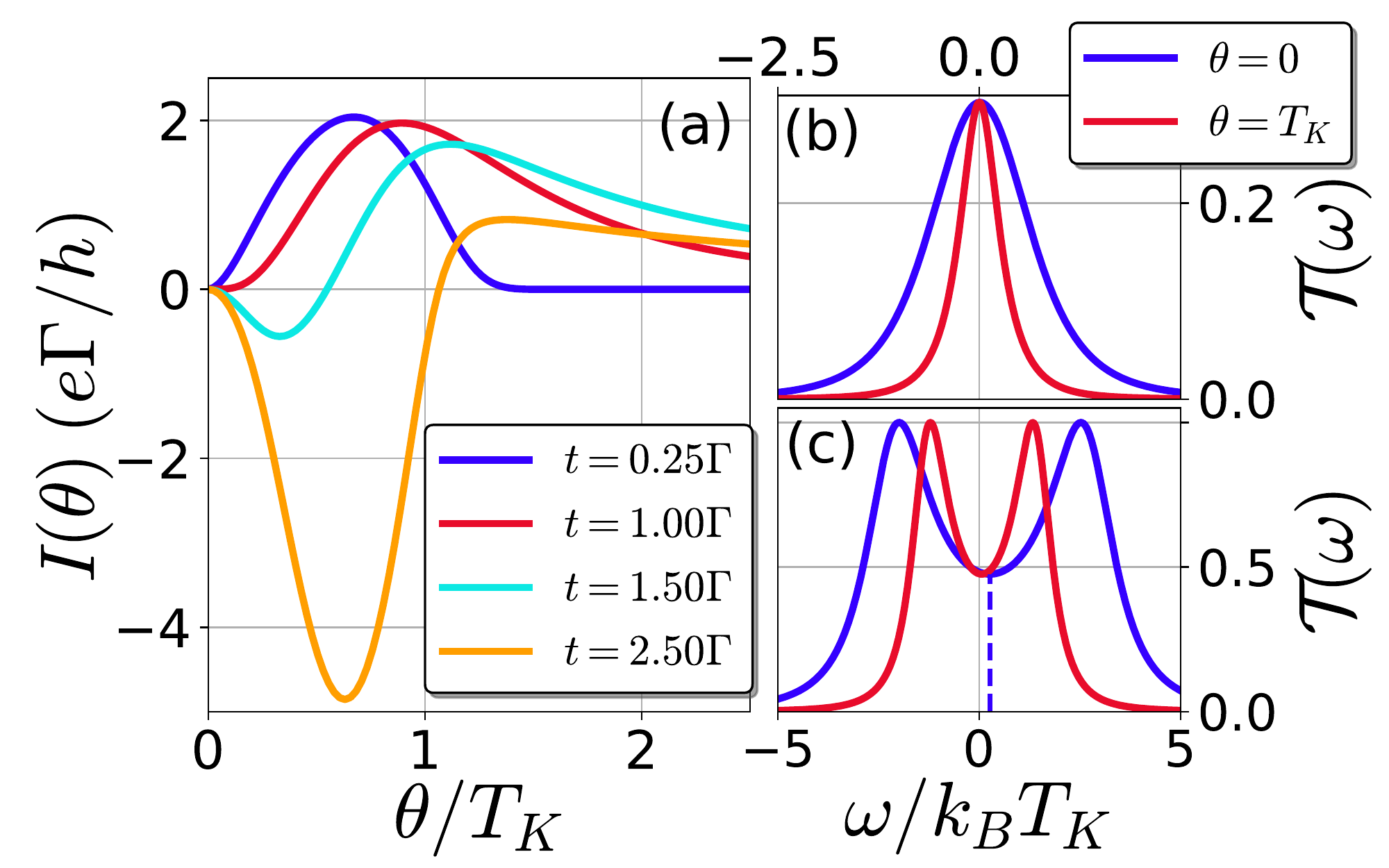}
\caption{(a) Thermoelectric current $I(\theta)$ for different values of the interdot tunnel coupling $t$. 
For clarity, all curves are rescaled with the factors: $10^{-6}$, $10^{-5}$, $5\times 10^{-5}$, and $10^{-4}$ for $t=0.25\Gamma$, $\Gamma$, $1.5\Gamma$, and $t=2.5\Gamma$, respectively.  
(b) and (c) show transmission probabilities $\calt(\omega)$ for $t/\Gamma < 1$ and $t/\Gamma > 1$, respectively.
Two different $\theta$ values are considered.
The dashed line in (c) indicates the position of local minimum showing that the transmission is not symmetric.
Same parameters as in Fig.~\ref{fig:1}.}
\label{fig:3}
\end{figure}

All previous features determine the characteristic behaviors of the heat $Q(\theta)$ and thermoelectric current $I(\theta)$. 
Figure~\ref{fig:2}(a) displays $Q(\theta)$ for different interdot tunnel couplings indicating the regimes (i) $t=0.25\Gamma$ , (ii) $t=\Gamma$ , and (iii) $t=1.5\Gamma, 2.5\Gamma$. 
$Q(\theta)$ is always positive as expected. We observe $Q(\theta)$ presents a maximum at a certain critical $\theta_c$. Then, a region of diminishing heat current appears for $\theta>\theta_c$ exhibiting a negative differential thermal conductance: the basis of a themal diode. Here, the DQD system allows the heat flow to occurs in one direction preferrentially \cite{Tdiode}.

In the linear regime ($\theta\rightarrow 0$) the heat flow increases linearly with $\theta$ according to the Fourier's law: $Q\approx K_0\theta$ with $K_0$ the linear thermal conductance which, at low temperatures follows $K_0\approx k_0 \mathcal{T}(\varepsilon_F)$ where $k_0$ is the thermal conductance quantum [see Fig.~\ref{fig:2}(c)] whereas for $\theta\rightarrow \infty$ the heat flow is suppressed since Kondo correlations at the hot reservoir are efficiently suppressed. More explicitly this result is observed in the nonlinear thermal conductance $K(\theta)=dQ/d\theta$ in Figure~\ref{fig:2}(b) where by increasing $\theta$ it makes $K(\theta)$ to change sign indicating that the DQD setup decouples due to a much smaller renormalized tunneling amplitude $\tilde{t}=t\tilde{b}_L \tilde{b}_R$.  As a consequence, we claim that the Wiedemann-Franz law is fulfilled as long as the system behaves as a Fermi liquid ($\theta<T_K$).

The thermoelectric transport of the system is characterized by the thermocurrent $I(\theta)$ depicted in Fig.~\ref{fig:3}(a). In contrast to the heat current, $I(\theta)$ shows distinctive differences depending on $t/\Gamma$. For $t/\Gamma < 1$, $I(\theta)$ is always positive, the transmission has a single peak centred at the positive frequency side, i.e., the thermocurrent has an electron-dominant character [see Fig.~\ref{fig:3}(b)]  \cite{Mig14,Mig17}. 
Increasing $\theta$ makes the peak of $\mathcal{T}(\omega)$ narrower, and eventually at larger $\theta$ such shrinking yields a suppressed $I(\theta)$.  When $t/\Gamma>1$ the transmission exhibits a double peak structure
due to the formation of coherent bonding and antibonding states. The transition from a singly-peak transmission when $t/\Gamma < 1$ to a double-peak transmission  for $t/\Gamma >1$  induces a sign change in $I(\theta)$ for some range of $\theta$ values [see Fig.~\ref{fig:3}(c)]. 
The explanation for such sign reversal in $I(\theta)$ is shown in Fig.~\ref{fig:3}(c) where the double dot transmission is plotted for  $t=1.5\Gamma$  at a small $\theta$ value. Notice that 
the transmission has more weight located at the negative frequency side in $\mathcal{T}(\omega)$ causing
a hole-like flow which gives negative $I(\theta)$.
By increasing $\theta\approx T_K$, the weight on the positive frequency side in $\mathcal{T}(\omega)$ starts to contribute adding an electron-like flow to $I(\theta)$.
Therefore, there is a temperature $\theta^*$ where the hole and electron contributions to $I(\theta)$ compensate each other, yielding a vanishing thermoelectric current.  
Eventually, when $\theta>\theta^*$ the weights on positive frequencies in $\mathcal{T}(\omega)$ dominate and therefore $I(\theta)>0$. Eventually at large $\theta$, $I(\theta)$ vanishes again since Kondo correlations on the hot dot are fully suppressed and hence left and right reservoirs become uncoupled. 

\emph{Results (II): Thermally driven Kondo-to-antiferromagnetic  crossover---}
In contrast with the results explained above, when $T_K$ becomes smaller the presence of $J$ is unavoidable, inducing the Kondo-to-AF transition. Therefore, we now include a non negligible AF interaction $J$ by adding one has to add within the MF approach
the term $\calh_J = J S_{L}\cdot S_{R}$ that depicts the antiferromagnetic interaction between localized spins $S_{L/R}$ with the coupling strength $J$ ($>0$) \cite{Geo99}.

\begin{figure}
\centering
\includegraphics[width=0.45\textwidth]{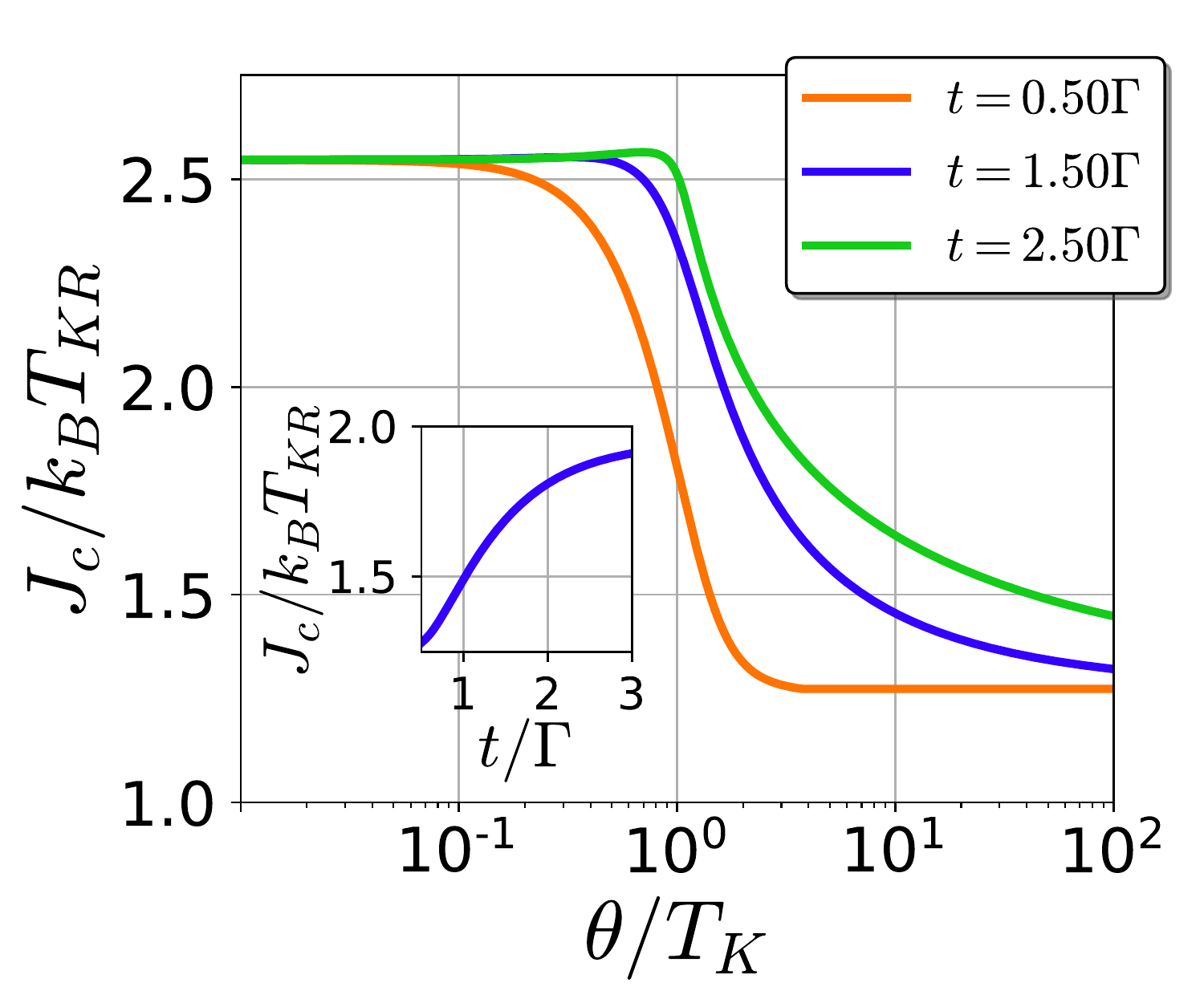}
\caption{$J_c/k_BT_{KR}$ for several values of the interdot tunnel coupling $t/\Gamma$. 
Inset: $J_c/k_BT_{KR}$ as a function of $t/\Gamma$ at $\theta = 3.1T_K$. 
Same parameters as in Fig.~\ref{fig:1}.}
\label{fig:4}
\end{figure}

When $J$ exceeds a critical value $J_c$, 
the localized dot spins favor the formation of an AF spin singlet state that blocks the charge transport. 
In serially coupled dots, several works have reported a crossover from the Kondo regime towards an AF single state between the dot spins \cite{Cra04,Chor12,Lar13}. 
Such an AF singlet state is manifested in the nonlinear conductance $\mathcal{G}$  exhibiting a splitting of value $\delta\approx 2J$ \cite{Aon01,Ros02,Pas05}.  
Besides, the critical value $J_c$ has been demonstrated to take the value \cite{Pas05} 
\begin{equation}\label{crossover}
J_c/T_{KR}\approx \frac{4}{\pi}\left (1+\frac{T_{KL}}{T_{KR}}\right)\,,
\end{equation}
where $T_{KR}>T_{KL}$. Previously, we showed that a thermal bias alters dramatically the behavior of the Kondo scales $T_{KL}$ ($\approx\tGamma_{L}$) and $T_{KR}$ ($\approx\tGamma_{R}$). 
Therefore, according to Eq.~\eqref{crossover}, $\theta$ affects the value of $J_c$. 
In Fig.~\ref{fig:4}, we depict $J_c/T_{KR}$ when $\theta$ is tuned for different $t/\Gamma$. 
The overall tendency is a transition from $J_c/T_{KR}\approx 8/\pi$ towards $J_c/T_{KR}\approx 4/\pi$ at large $\theta$ due to the suppression of the Kondo correlations in the hot dot $T_{KL}\approx 0$. 
The inset of Fig.~\ref{fig:4} displays the behavior of the critical value $J_c/T_{KR}$ for a fixed $\theta=3.25T_K$  as a function of $t/\Gamma$. 
$J_c/T_{KR}$ increases with $t/\Gamma$ because Kondo correlations (for low/moderate $\theta$ values) are reinforced when the interdot tunneling coupling $t/\Gamma$ increases.
Consequently, a stronger AF coupling is needed to suppress Kondo correlations. 

\emph{Closing--}
Summarizing, the two-impurity Kondo system driven by a thermal bias $\theta$ has been examined. 
When Kondo correlations are dominant ($J=0$), $\theta$ has a strong influence on the Kondo scales and consequently on the transport properties of the Kondo-based setups. Our findings indicate that for sufficiently large $\theta$ the Kondo correlations 
are destroyed for the dot coupled to the hot reservoir regardless the interdot strength coupling. Since the interdot coupling is renormalized by the Kondo correlations for both dots the DQD setup gets decoupled leading to suppressed electrical and heat flows. For non neglibible antiferromagnetic spin exchange coupling, we report the influence of $\theta$ on the Kondo-to-AF crossover.  For a finite $J$, we find that such crossover takes place around a critical value that for a small $\theta$ takes the value $J_c/T_{KR} = 8/\pi$ whereas for a large $\theta$, $J_c$ reaches the value of $\approx4/\pi$. This result is due to the quenching Kondo correlations on the hot reservoir when $\theta$ augments. Finally, we highlight that all our observations might be experimentally tested due to the huge progress on thermoelectrical transport through nanostructures. 

\emph{Acknowledgments---}
We are grateful to D. S{\'a}nchez for helpful discussions. 
This work was supported by the MINECO grants No.~FIS2014-52564 and MAT2017-82639 and a PhD grant from CAIB.

\end{document}